\documentclass[11pt,a4paper]{article}
\usepackage{amssymb,amsmath,fullpage}
\usepackage[round]{natbib}
\usepackage{amsfonts,amsbsy}
\usepackage{graphicx}
\usepackage{bm}
\usepackage{lineno}

\newcommand*\patchAmsMathEnvironmentForLineno[1]{%
  \expandafter\let\csname old#1\expandafter\endcsname\csname #1\endcsname
  \expandafter\let\csname oldend#1\expandafter\endcsname\csname end#1\endcsname
  \renewenvironment{#1}%
     {\linenomath\csname old#1\endcsname}%
     {\csname oldend#1\endcsname\endlinenomath}}%
\newcommand*\patchBothAmsMathEnvironmentsForLineno[1]{%
  \patchAmsMathEnvironmentForLineno{#1}%
  \patchAmsMathEnvironmentForLineno{#1*}}%
\AtBeginDocument{%
\patchBothAmsMathEnvironmentsForLineno{equation}%
\patchBothAmsMathEnvironmentsForLineno{align}%
\patchBothAmsMathEnvironmentsForLineno{flalign}%
\patchBothAmsMathEnvironmentsForLineno{alignat}%
\patchBothAmsMathEnvironmentsForLineno{gather}%
\patchBothAmsMathEnvironmentsForLineno{multline}%
}

\renewcommand{\theequation}{\thesection.\arabic{equation}}
\let\ssection=\section
\renewcommand{\section}{\setcounter{equation}{0}\ssection}

\def\e{\mathrm{e}}
\def\i{\mathrm{i}}
\def\d{\mathrm{d}}

\def\beq{\begin{equation}}
\def\eeq{\end{equation}}

\def\eps{\varepsilon}

\newcommand{\eqn}[1]{(\ref{eqn:#1})}
\newcommand{\lab}[1]{\label{eqn:#1}}
\newcommand{\inter}[1]{\quad \textrm{#1} \quad}

\newcommand{\hs}[1]{{{h}}^{(#1)}}
\newcommand{\Hs}[1]{{{H}}^{(#1)}}
\newcommand{\ms}[1]{{{m}}^{(#1)}}

\newcommand\lap{\nabla^2}

\def\XXint#1#2#3{{\setbox0=\hbox{$#1{#2#3}{\int}$}
\vcenter{\hbox{$#2#3$}}\kern-.5\wd0}}

\title{Coastal imbalance: generation of oceanic Kelvin waves by atmospheric perturbations}
\author{Jacques Vanneste}

\date{\small School of Mathematics and Maxwell Institute for Mathematical Sciences, \\
University of Edinburgh, Edinburgh EH9 3FD, UK}

\begin{document}
\maketitle

\begin{abstract}
\noindent
The response of a semi-infinite ocean to a slowly travelling atmospheric perturbation crossing the coast provides a simple example of the breakdown of nearly geostrophic balance induced by a boundary. We examine this response in the linear shallow-water model at small Rossby number $\eps \ll 1$. Using matched asymptotics we show that a long Kelvin wave, with $O(\eps^{-1})$ length scale and $O(\eps)$ amplitude relative to quasigeostrophic response, is generated as the perturbation crosses the coast. Accounting for this Kelvin wave restores the conservation of mass which is violated in the quasigeostrophic approximation.

\end{abstract}

\section{Introduction}

This paper is motivated by fundamental aspects of quasigeostrophic (QG) and higher-order balances in the ocean and, specifically, their accuracy in the presence of boundaries. It is well understood that, in unbounded or periodic domains and with smooth forcing, a suitable initialisation can filter out inertia-gravity waves up to high accuracy \citep{warn-et-al}. Initial conditions chosen to satisfy a balance relation defined perturbatively to $O(\eps^n)$, where $\eps \ll 1$ is the Rossby number, can develop inertia-gravity waves with amplitudes that are at most $O(\eps^{n+1})$. Optimal-truncation arguments then indicate that inertia-gravity waves that cannot be filtered by even the best initialisation -- in other words  spontaneously generated inertia-gravity waves -- have amplitudes that are exponentially small in $\eps$ \citep{v08,v13}. It is less widely appreciated that the presence of a horizontal boundary in the form of a long coastline drastically changes this state of affairs. Long Kelvin waves that propagate along the coastline have low frequencies  matching the frequencies of the balanced motion \citep[e.g.][]{zeit18}. As a result, such long Kelvin waves are generated spontaneously, typically with $O(\eps)$ amplitudes,  by well balanced flows through a process analogous to Lighthill radiation.

This is not a new observation. \citet{doro-lari} show that the reflection of a linear shallow-water Rossby wave on an infinite wall is accompanied by the emission of a Kelvin wave with $O(\eps)$ amplitude. This wave emission resolves an inconsistency of the QG approximation highlighted by the authors, namely its failure to conserve total mass and circulation along the wall. 
In their comprehensive study of geostrophic adjustment in the presence of an infinite wall, \cite{rezn-grim}  predict the generation of Kelvin waves by arbitrary localised geostrophic motion and relate it to mass and circulation conservation.  The topic is further examined by \citet{rezn-suty} who clarify the role played by the size of the domain and the differences between localised and periodic flows. 

The aim of this paper is to present a straightforward example of Kelvin-wave generation by geostrophic motion in the presence of a long coast modelled as an infinite wall. In this example, a slowly travelling wind stress induces an ocean response that is nearly geostrophic at all times but is accompanied by the transient emission of a Kelvin wave as the wind stress crosses the coast.  
The simplicity of the configuration and model (linear shallow water, \S\ref{sec:model}) enables us to obtain completely explicit results for $\eps \ll 1$, including for the form of the Kelvin wave, which is derived using matched asymptotics (\S\ref{sec:asympt}). Numerical solutions of the linear shallow-water equations confirm these results (\S\ref{sec:numerics}). 

 In addition to illustrating a fundamental feature of near-geostrophic balance, namely its inevitable breakdown at algebraic order in $\eps$ caused by boundaries, the problem studied has a practical interest: coastally-trapped waves (Kelvin, shelf and edge waves) generated by weather systems are well documented and have been extensively studied \citep{kaji62,thom70,gill-schu,grim88b,tang-grim}. Their most spectacular manifestation is as part of storm surges caused by hurricane landfalls \citep[e.g.][]{yank09}. Our results provide an elementary demonstration of the mechanism underpinning this wave generation, albeit with an assumption of small Rossby number.

\section{Model} \label{sec:model}

\begin{figure}
\begin{center}
\includegraphics[height=6cm]{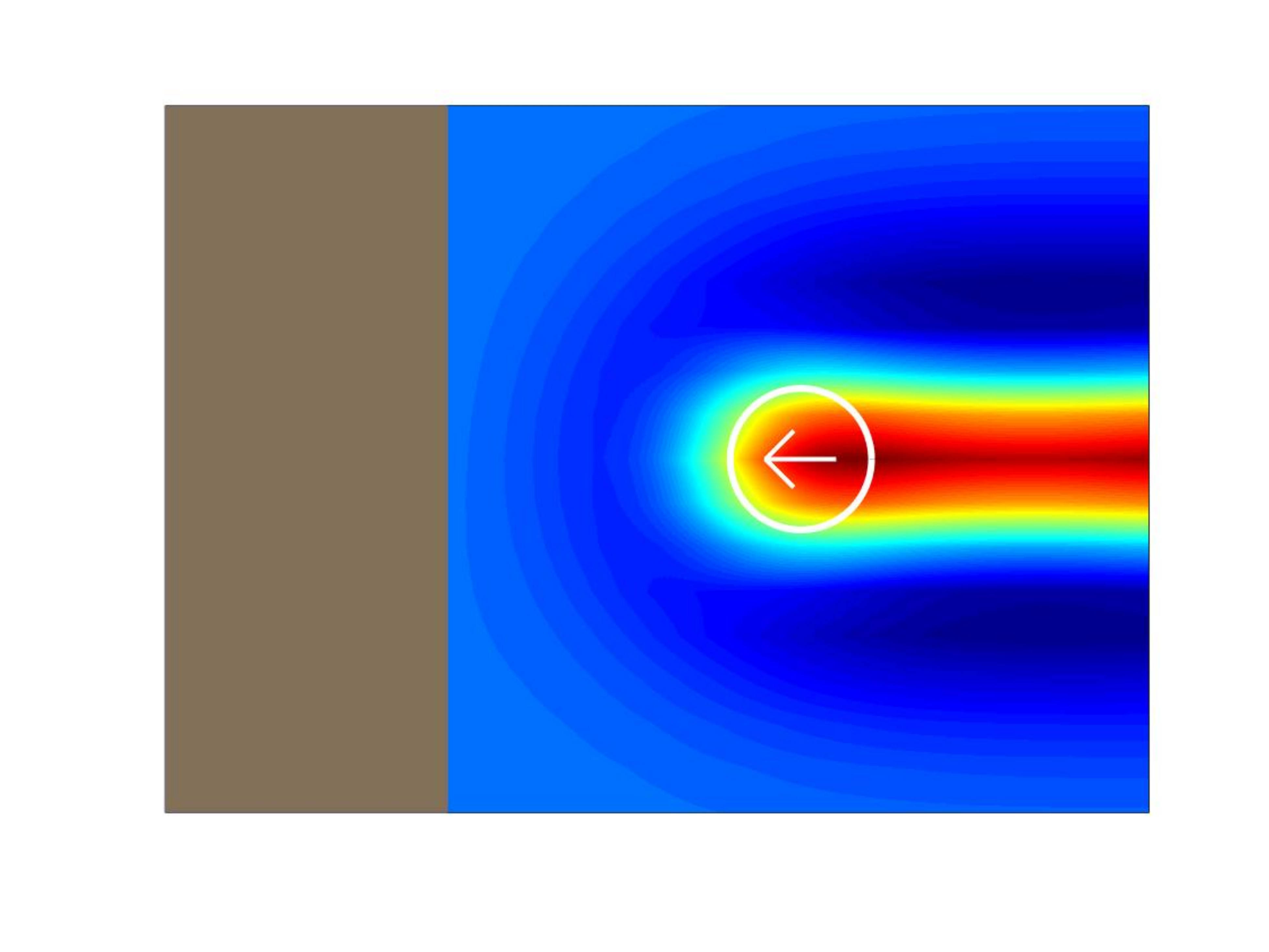}
\caption{Schematic of the model: an atmospheric perturbation indicated by the white circle travels westward over the ocean, leaving a geostrophically balanced flow in its wake (the height field $h$ is shown by the colour scale), before crossing the coast and continuing over the land (shown in brown). The case of an eastward-travelling perturbation starting overland is also considered.}
\label{fig:coastSchematic}
\end{center}
\end{figure}

We examine the response of the ocean to an atmospheric perturbation (cyclone or anti-cyclone) as this crosses the coast. 
We model this process using the linear rotating shallow-water equations forced by a skew-gradient stress. In dimensionless form, the governing equations are
\begin{subequations} \lab{SWE}
\begin{align}
\eps u_t - v &= - h_x - \eps \Phi_y, \lab{momu} \\
\eps v_t + u &= - h_y + \eps \Phi_x, \lab{momv} \\
\eps \lambda^2 h_t + u_x + v_y &=  0. \lab{cont}
\end{align}
\end{subequations}
Here $\eps = U/(fL)$ is the Rossby number, defined in terms of the inertial frequency $f>0$ and of the velocity $U$ and lengthscale $L$ of the perturbation, and $\lambda^2 = L^2/L_D^2 = f^2 L^2/(gH)$ is an inverse Burger number \citep[e.g.][\S5.1]{vall17}. The domain is the half plane $x \ge 0$. The boundary condition 
\beq
u=0 \quad \textrm{at} \ \ x = 0
\lab{bcu}
\eeq
applies along the coast. 

The forcing is proportional to $\eps$ -- so the ocean response is primarily geostrophic -- and defined by the $O(1)$ scalar function $\Phi$ taken as a localised function of $x \mp t$ and $y$. The upper sign corresponds to a perturbation that travels westward over the ocean for $t<0$, makes landfall at $t=0$ and continues overland for $t>0$; the lower sign corresponds to a perturbation travelling eastward overland for $t<0$, crossing the coast at $t=0$ and continuing over the ocean for $t>0$.  Figure \ref{fig:coastSchematic} illustrates the case of the westward-travelling perturbation. 
To fix ideas, and  for the numerical simulations below, we take $\Phi$ to be the Gaussian
\beq
\Phi(x \mp t,y) = (2\pi)^{-1} \e^{-\left((x\mp t)^2+y^2\right)/2}.
\lab{gaussian}
\eeq
This corresponds to an anti-cyclonic wind stress, which raises the sea surface  in its wake. 
We consider the evolution from a state of rest as $t \to -\infty$. This eliminates any wave generation by adjustement and implies that $u, \, v,\, h \to 0$ as $y \to \pm \infty$.


We can eliminate $u$ and $v$ from \eqn{SWE} to obtain the single equation
\beq
(\lap - \lambda^2) h_t - \eps^2 \lambda^2 h_{ttt} = \lap \Phi,
\lab{heqn}
\eeq
with $\lap = \partial_x^2 + \partial_y^2$ the Laplacian. The associated boundary condition is found by combining \eqn{momu}, \eqn{momv} and \eqn{bcu} to obtain
\beq
h_y + \eps h_{xt}= \eps \Phi_x - \eps^2 \Phi_{yt} \quad \textrm{at} \ \ x = 0.
\lab{hbc}
\eeq
Note that \eqn{heqn}--\eqn{hbc} are consistent with the conservation of the total mass
\beq
m(t) =  \int \!\! \int_{x \ge 0} h \, \d x \d y
\eeq
which follows from \eqn{cont}--\eqn{bcu}: integrating \eqn{heqn} and taking \eqn{hbc} into account gives
\beq
m_t + \eps^2 m_{ttt} = \lambda^{-2} \int_{x=0} \left(\Phi_x - h_{xt} \right) \, \d y = \lambda^{-2} \int_{x=0} \left(\eps^{-1} h_y + \eps \Phi_{yt} \right) \, \d y = 0
\eeq
since $h, \, \Phi_t \to 0$ as $y \to \pm \infty$.

\section{Small-Rossby asymptotics} \label{sec:asympt}

We now solve \eqn{heqn}--\eqn{hbc} in the small-Rossby-number regime $\eps \ll 1$ with $\lambda=O(1)$ corresponding to  standard QG scaling. Expanding
\beq
h = \hs{0} + \eps \hs{1} + \cdots
\eeq
and introducing into \eqn{heqn}--\eqn{hbc} gives, at leading order, the linearised QG equation
\beq
(\lap - \lambda^2) \hs{0}_t = \lap \Phi
\lab{qg}
\eeq
with  boundary condition
\beq
\hs{0}_y = 0 \quad \textrm{at} \ \ x = 0.
\lab{h0bc0}
\eeq
Since $h \to 0$ as $y \to \pm \infty$, this implies
\beq
\hs{0} = 0 \quad \textrm{at} \ \ x = 0.
\lab{h0bc}
\eeq
As \citet{doro-lari}, \citet{rezn-grim} and \citet{rezn-suty} point out, this boundary condition is inconsistent with the conservation of the total mass 
associated with $\hs{0}$. Indeed integrating \eqn{qg} gives
\beq
\ms{0}_t= \lambda^{-2} \int_{x=0} \left( \Phi_x - \hs{0}_{xt} \right) \, \d y,
\lab{m0t}
\eeq
which is not constrained to vanish by \eqn{h0bc}. This is typical of QG dynamics in large domains, with boundary lengths that are $O(\eps^{-1})$ or larger compared with the deformation radius. In contrast, for $O(1)$ boundary lengths as often assumed implicitly, \eqn{h0bc0} implies only that $\hs{0} = C(t)$ on the boundary, with a function $C(t)$ determined to ensure total mass conservation. See \citet{rezn-suty} for a detailed discussion of the two regimes. Note that the non-vanishing of \eqn{m0t} also implies a failure of
QG to represent correctly the evolution of the circulation along the coast.  

Mass conservation is resolved by considering the correction $\eps \hs{1}$ to $\hs{0}$. At $O(\eps)$, \eqn{heqn}--\eqn{hbc} give
\begin{subequations}\lab{h1}
\beq 
(\lap - \lambda^2) \hs{1}_t = 0 \inter{with} \hs{1}_y = \Phi_x - \hs{0}_{xt} \quad \textrm{at} \ \  x=0. \tag{\theequation a,b}
\eeq
\end{subequations}
These equations can be solved, but not in a way that ensures $\hs{1} \to 0$ as $y \to \pm \infty$ since (\ref{eqn:h1}b) implies the jump
\beq \lab{jumph1}
\left[ \hs{1}(0,y,t) \right]_{y \to -\infty}^{y \to \infty} = \int_{x=0} \left( \Phi_x - \hs{0}_{xt} \right) \, \d y = \lambda^2 \ms{0}_t.
\eeq
The difficulty arises because (\ref{eqn:h1}b) is not uniformly valid as $y \to \pm \infty$. The solution $\hs{1}$ to \eqn{h1} should be regarded as an inner solution, to be matched to an outer solution, $\Hs{1}(x,Y,t)$ say, valid in the region $Y=\eps y = O(1)$.
This outer solution is found from \eqn{heqn}--\eqn{hbc} to satisfy
\begin{subequations} \lab{Hs}
\beq 
\Hs{1}_{xxt} - \lambda^2 \Hs{1}_t = 0 \inter{with} \Hs{1}_Y + \Hs{1}_{xt} = 0  \quad \textrm{at} \ \ x= 0, \tag{\theequation a,b}
\eeq
\end{subequations}
since $\Phi$ is exponentially small for $Y=O(1)$, together with matching conditions as $Y \to 0^\pm$. It takes the form  
\beq
\Hs{1}(x,Y,t) = K(Y,t) \, \e^{- \lambda x},
\eeq
as required by (\ref{eqn:Hs}a) and the condition $\Hs{1} \to 0$ as $x \to \infty$, and corresponds to a Kelvin wave. From (\ref{eqn:Hs}b) the amplitude $K(Y,t)$ satisfies $ K_Y - \lambda K_t = 0$, with a jump across $Y=0$ that matches \eqn{jumph1}. This can be rewritten compactly as 
\beq
K_Y - \lambda K_t = \lambda^2 \ms{0}_t \, \delta(Y),
\lab{KW}
\eeq 
with $\delta(Y)$ the Dirac distribution, and solved to obtain
\beq
K(Y,t) = - \lambda^2 \ms{0}_t(t + \lambda Y) \, \Theta(-Y),
\lab{KWsol}
\eeq
where $\Theta(Y)$ denotes the Heaviside step function. This represents a long Kelvin wave, propagating with the coast to its right (here southward) from the point where the atmospheric perturbation makes contact with the coast. The large Kelvin-wave speed $-(\eps \lambda)^{-1}$ ensures that, although the wave has a small $O(\eps)$ amplitude, it carries an $O(1)$ mass, thus balancing the change of the mass $\ms{0}$ associated with the leading-order QG solution. We verify that $\ms{0} + \eps \ms{1} = \mathrm{const}$ by computing 
\beq
\eps \ms{1}_t = \eps \int\!\!\int_{x \ge 0} \hs{1}_t \, \d x \d y =  \int\!\!\int_{x \ge 0} \Hs{1}_t \, \d x \d Y
= \lambda^{-1} \int K_t \, \d Y = - \ms{0}_t,
\eeq 
where the last equality follows from integrating \eqn{KW} in $Y$.

\section{Gaussian perturbation} \label{sec:numerics}

We now obtain explicit predictions and compare them with the results of numerical simulations of the linear shallow-water equations \eqn{SWE} in the case of a  Gaussian perturbation \eqn{gaussian}. We solve the QG equation \eqn{qg}  for $\hs{0}$ with boundary condition \eqn{h0bc} by writing
\beq
\Phi(x\mp t,y) = \iint \hat \Phi(k,l) \e^{\i(k (x \mp t) + ly)} \, \d k \d l \ \ \textrm{and} \ \  
\hs{0}_t(x,y,t) = \iint \hat g(x,k,l) \e^{\i(\mp k t + ly)} \, \d k \d l,
\lab{ft}
\eeq
where $\hat \Phi(k,l)=(2\pi)^{-2} \e^{-(k^2+l^2)/2}$ and the function $\hat g$ is to be determined. Introducing \eqn{ft} into \eqn{qg} yields
\beq
\hat g_{xx} - (l^2 + \lambda^2) \hat g = -(k^2 + l^2) \hat \Phi(k,l) \e^{\i k x}.
\eeq
Solving and imposing that $\hat g(x=0,k,l)=0$ as required by \eqn{h0bc} leads to
\beq
\hat g(x,k,l) = \frac{k^2+l^2}{k^2 + l^2 + \lambda^2} \left(\e^{\i k x} - \e^{-\sqrt{l^2+\lambda^2} x} \right) \hat{\Phi}(k,l).
\lab{ghat}
\eeq
This gives a Fourier representation for $\hs{0}_t$ and, by integration in time from $t \to -\infty$, $\hs{0}$. 

\begin{figure}
\begin{center}
\includegraphics[height=6cm]{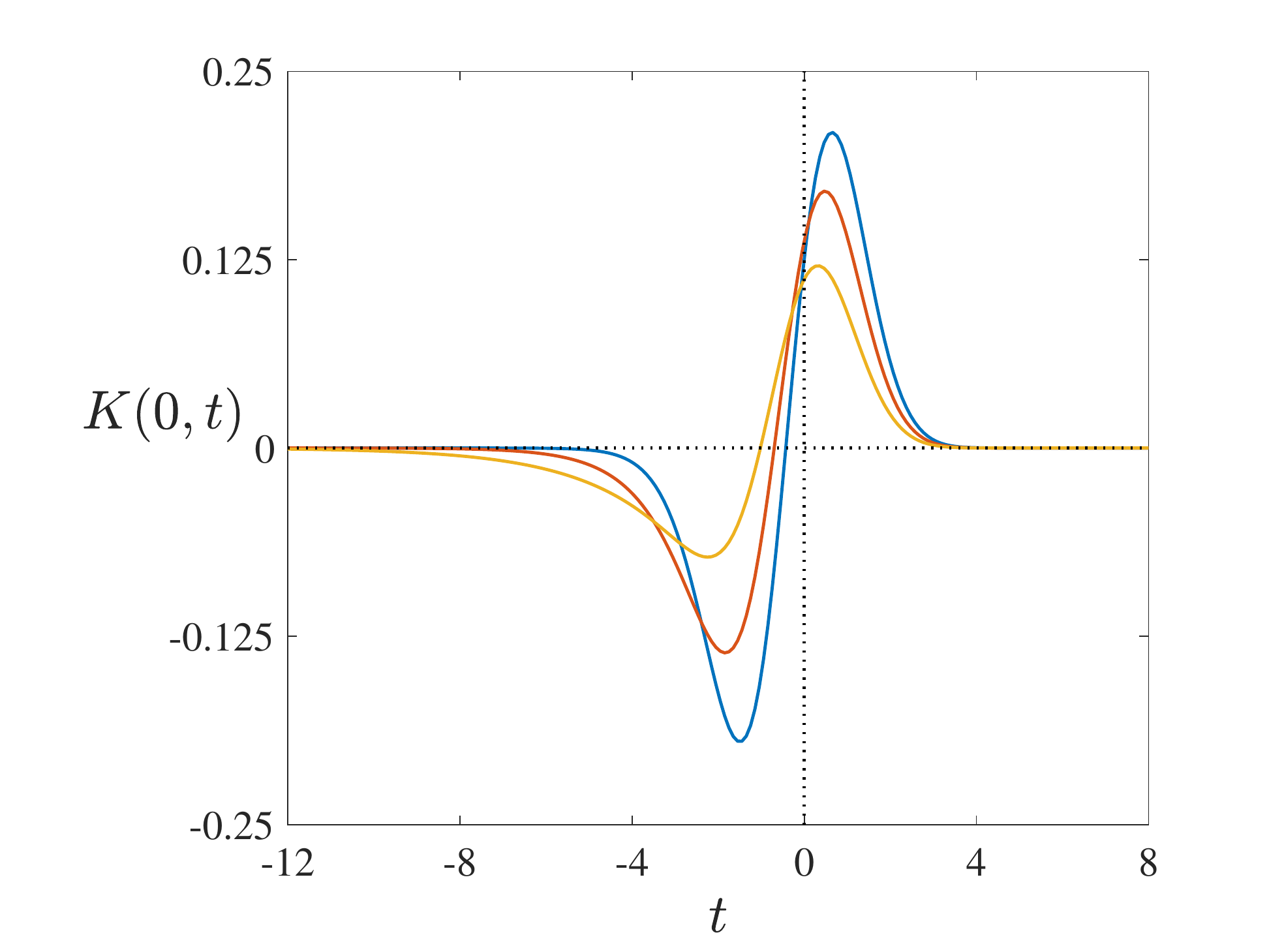}
\caption{Evolution of the height at $(x,y)=0$, where the atmospheric perturbation crosses the coast, according to the outer solution \eqn{hc} for $\lambda=0.5$ (orange), $1$ (red) and $2$ (blue).}
\label{fig:waveMaker}
\end{center}
\end{figure}

We focus on the Kelvin-wave response which, according to \eqn{KWsol} is determined by the mass rate of change $\ms{0}_t$. We compute  $\ms{0}_t$ from its definition \eqn{m0t}, using \eqn{ft}, \eqn{ghat} and $\int \e^{ily} \, \d y = 2 \pi \delta(l)$ to find
\beq
\ms{0}_t = 2 \pi \int \frac{\i \lambda k - k^2}{\lambda(k^2 + \lambda^2)} \hat \Phi(k,0) \e^{\mp \i k t} \, \d k.
\lab{ms0}
\eeq
Introducing this into \eqn{KWsol} gives an explicit form for the Kelvin wave generated by the atmospheric perturbation. In particular, at the coast, for $y<0$  and away from the inner region $y=O(1)$,
\beq \lab{hc}
h(0,y,t) \sim \eps K(\eps y,t) =- 2 \pi \eps \int \frac{\i \lambda^2 k - \lambda k^2}{k^2 + \lambda^2}  \e^{\mp \i k (t + \eps \lambda y)} \hat \Phi(k,0) \, \d k + O(\eps^2),
\eeq
since $\hs{0}=0$ for $x=0$. Figure \ref{fig:waveMaker} shows $K(0,t) = - \lambda^2 \ms{0}_t$ as a function of $t$ for $\lambda=0.5,\, 1,\, 2$ and for an atmospheric perturbation travelling westward, from the open ocean towards the coast (i.e.\ with the factor $\e^{\i k t}$ in \eqn{ms0}--\eqn{hc}). This function approximates the evolution of the height at the location of the landfall as described by the outer solution $\Hs{1}$; it can be interpreted as the amplitude of a wavemaker generating the Kelvin-wave response to the atmospheric perturbation. Note the asymmetry of the evolution about $t=0$, with a depression phase ($t<0$) that is weaker and lasts longer than the elevation phase ($t>0$). This asymmetry is constrained by the vanishing of the integral of $K(0,t)$ for $t \in (-\infty,\infty)$ which reflects the fact that the change in the QG mass $\ms{0}$ is purely transient with the forcing chosen.
The asymmetry increases as $\lambda$ decreases. The case of an atmospheric perturbation travelling eastward, from overland towards the ocean, is deduced by reversing the sign of $t$.

We illustrate and verify the above predictions by carrying out numerical simulations of the linear shallow-water equations \eqn{SWE}. The numerical model used discretises the fields $(u,v,h)$ on a staggered grid in the $x$-direction, with $u$ represented at grid points $j \Delta x$ for $j=1,2,\cdots$ and $v$ and $h$ at grid points $(j-1/2)\Delta x$. A Fourier expansion is used in the $y$-direction with an assumption of periodicity. The $x$-derivatives are approximated by central differences; the $y$-derivatives are computed spectrally. The domain size is $10 \times 160$, with the $y=0$ axis, corresponding to the path of the  
atmospheric perturbation, placed at a distance 40 of the northern boundary of the domain. These choices ensure that the distance from the path of the perturbation to the southern boundary is large enough for the long Kelvin wave to propagate south along the coast without being affected by the $y$-periodicity of the domain. 
The physical parameters for the simulations reported are $\lambda=1$ and $\eps=0.1$; the numerical parameters are: 128 grid points in $x$, 256 Fourier modes in $y$, and a time step $\Delta t = 0.1 \eps \Delta x$ with $\Delta x = 10/128$.

\begin{figure}
\begin{center}
\hspace{-1.1cm} \includegraphics[height=8.2cm]{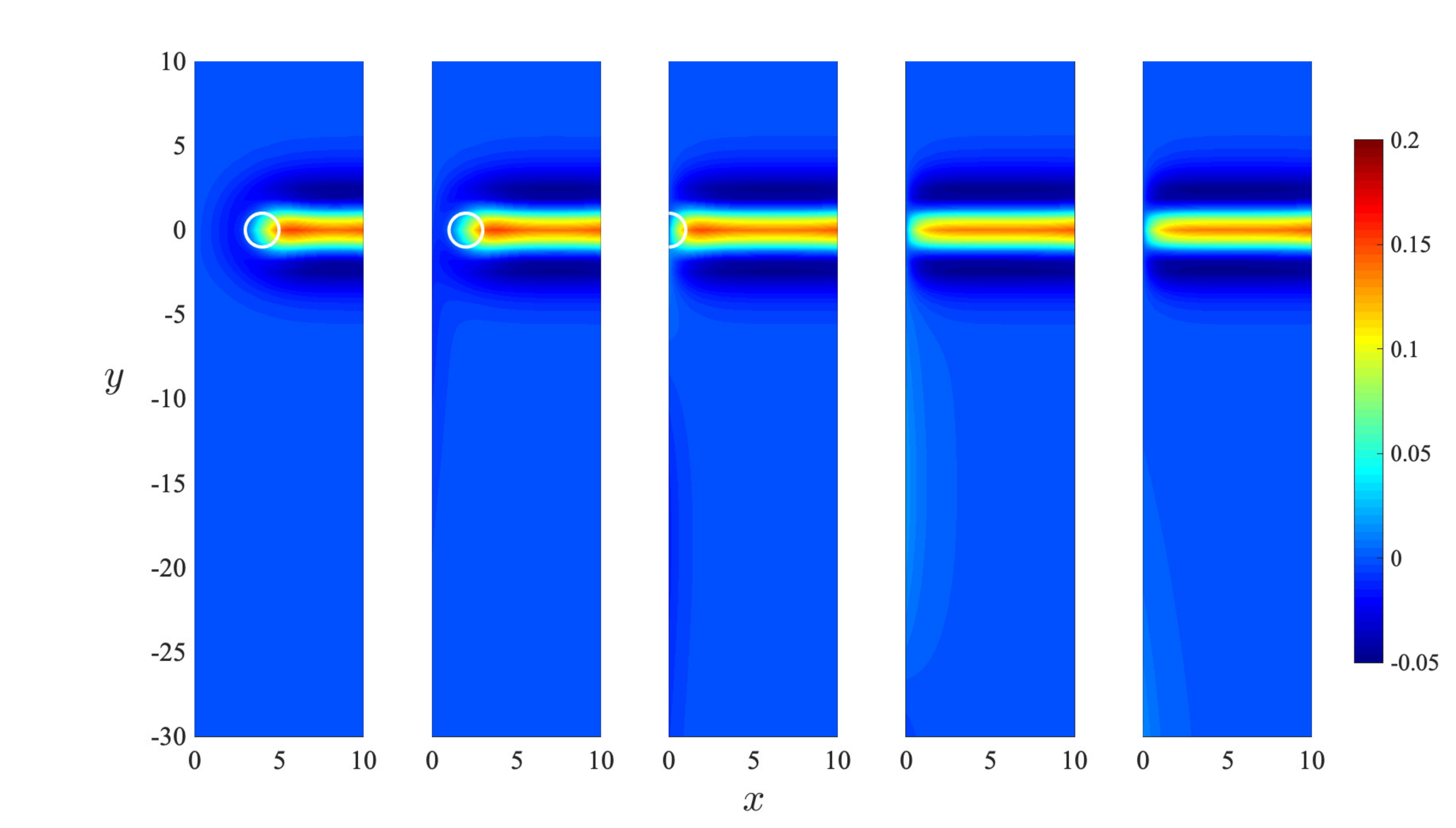}
\caption{Height field $h$ for a Gaussian atmospheric perturbation propagating westward from the open ocean towards the coast, shown at times $t=-4,\, -2, \, 0 \textrm{ (landfall time)},\, 2$ and $4$, from left to right. The parameters are $\lambda=1$ and $\eps=0.1$. The location of the perturbation is indicated in the first three panels by the white circle with radius $1$ corresponding to the length scale of Gaussian.}
\label{fig:hAllWestward}
\end{center}
\end{figure}

Figure \ref{fig:hAllWestward} shows successive snapshots of the height field $h$ for a westward-travelling perturbation as this makes landfall. At $t=-4$, when the centre of the perturbation is a distance 4 away from the coast (recall that the speed of the perturbation is $1$ in dimensionless units), the response of the ocean is well balanced, QG to a good approximation and, in particular, symmetric about the $x$-axis as the QG solution \eqn{ft}--\eqn{ghat} indicates. 
At $t=-2$, there is a weak signal of negative $h$ along the coast and south of the atmospheric perturbation, breaking the $\pm y$ symmetry. 
This is the signature of the $O(\eps)$ Kelvin wave generated by the mass imbalance associated with the QG response.
This initial depression wave propagates south, with its maximum reaching $y=-20$ at $t=0$. It is followed by the propagation of an elevation visible for $t=2$ and $4$.  

\begin{figure}
\begin{center}
\includegraphics[height=5cm]{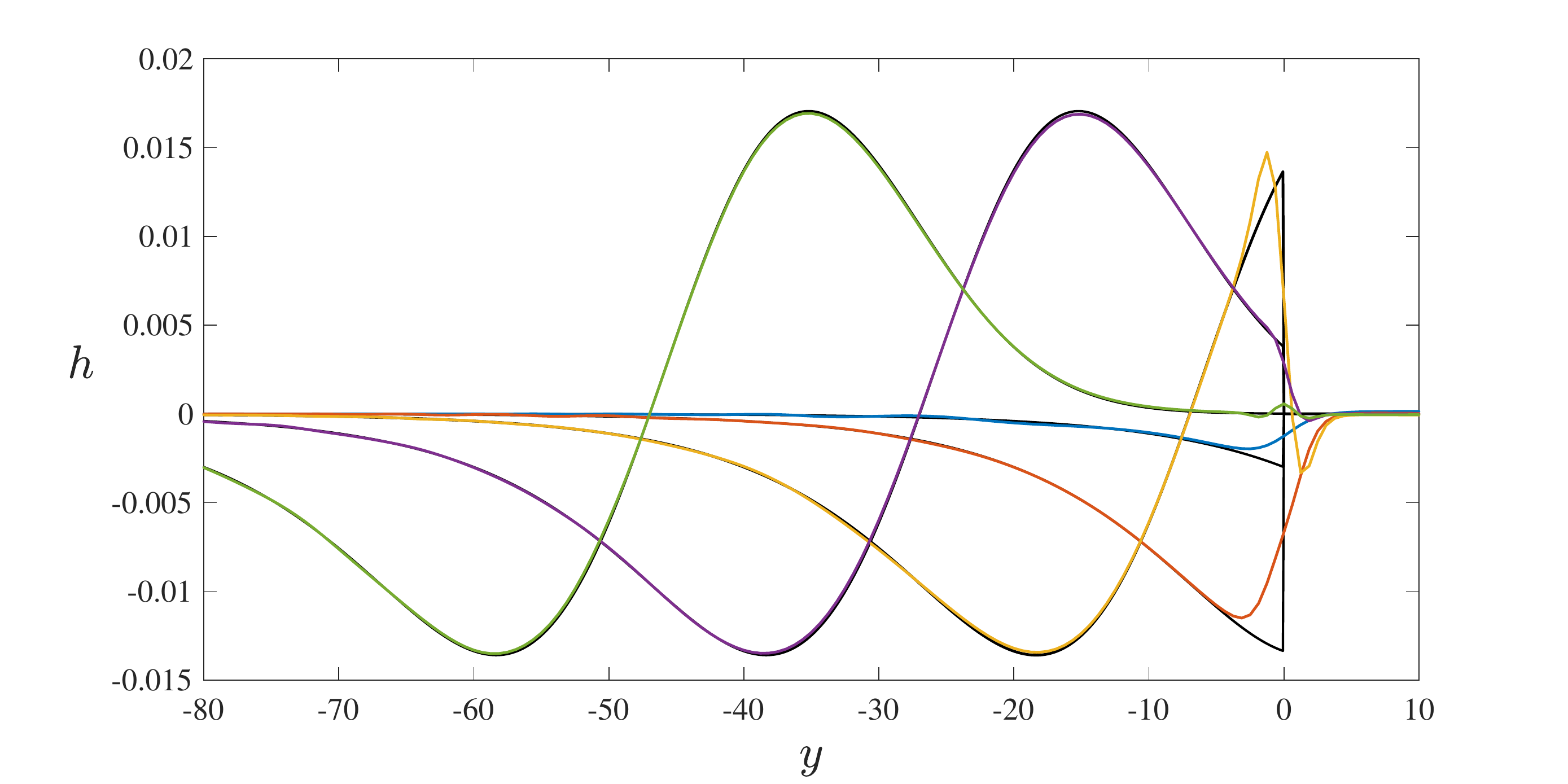}
\caption{Height field along the coast $x=0$ at times $t=-4$ (blue), $-2$ (red), $0$ (landfall time, orange), $2$ (purple) and $4$ (green) for the simulation in figure \ref{fig:hAllWestward}, showing the generation near $y=0$ and southward propagation of a Kelvin wave. The predictions based on the matched-asymptotics result \eqn{hc} are shown by the black curves closely matching the coloured curves except near $y=0$.}
\label{fig:hCoastWestward}
\end{center}
\end{figure}

A clearer depiction of the Kelvin-wave propagation is given in figure \ref{fig:hCoastWestward} which shows the height field along the coast, where the QG contribution $\hs{0}$ vanishes. The figure confirms the validity of the matched-asymptotics prediction \eqn{hc} which approximates the numerical solution with remarkable accuracy except, of course, in the `inner' region $y=O(1)$ where \eqn{h1} applies. Thus the simple picture of a Kelvin wave generated by a wavemaker with time dependence as shown in figure \ref{fig:waveMaker} is well justified.

\begin{figure}
\begin{center}
\hspace{-1.1cm} \includegraphics[height=8.2cm]{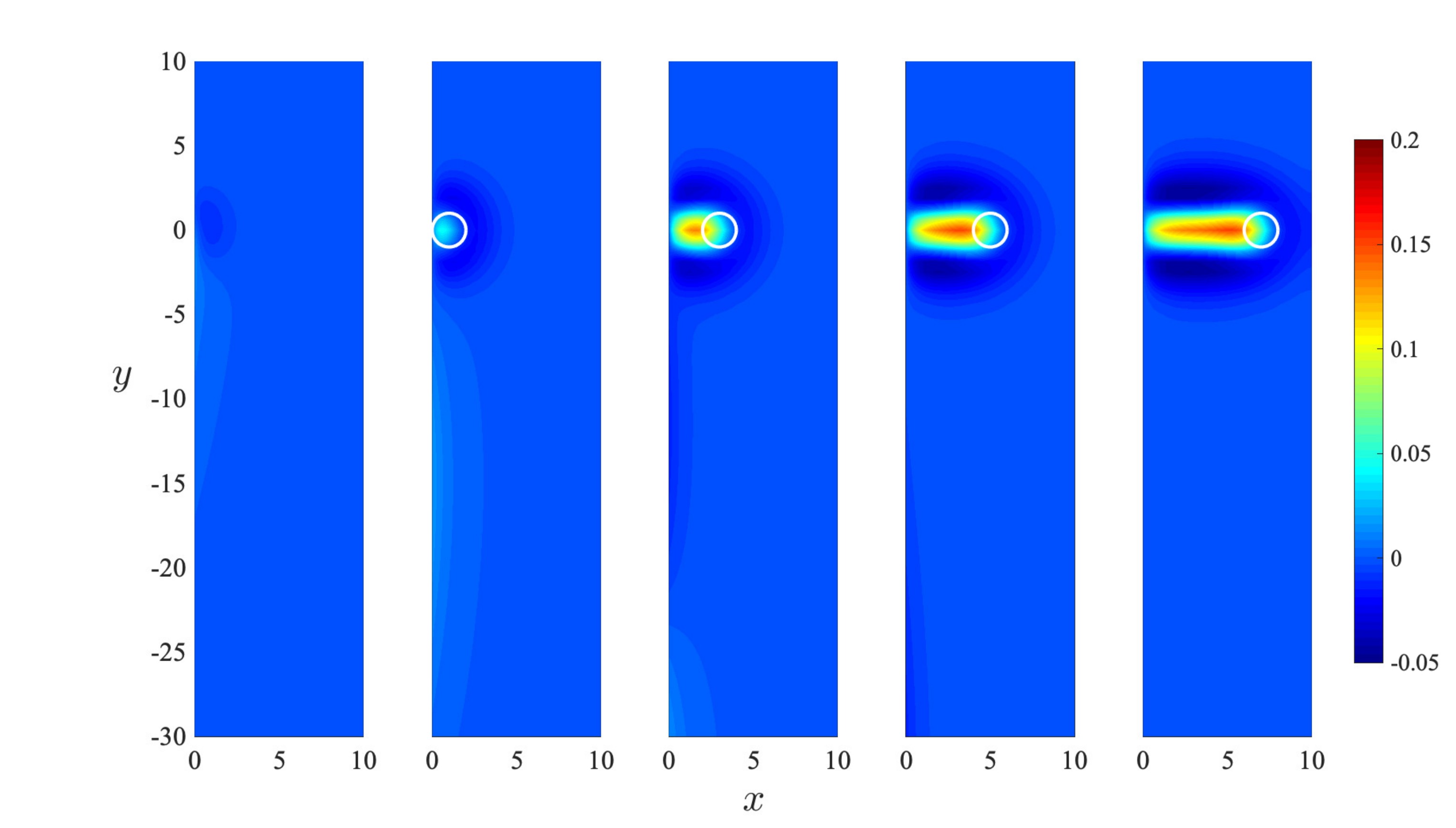}
\caption{Same as figure \ref{fig:hAllWestward} but for an atmospheric perturbation travelling eastward, from overland to the open ocean, crossing the coast at $t=0$. The height field $h$ is shown at times $t=-1,\, 1,\, 3,\, 5$ and $7$ from left to right. The location of the perturbation is indicated by the white circles.} 
\label{fig:hAllEastward}
\end{center}
\end{figure}

\begin{figure}
\begin{center}
\includegraphics[height=5cm]{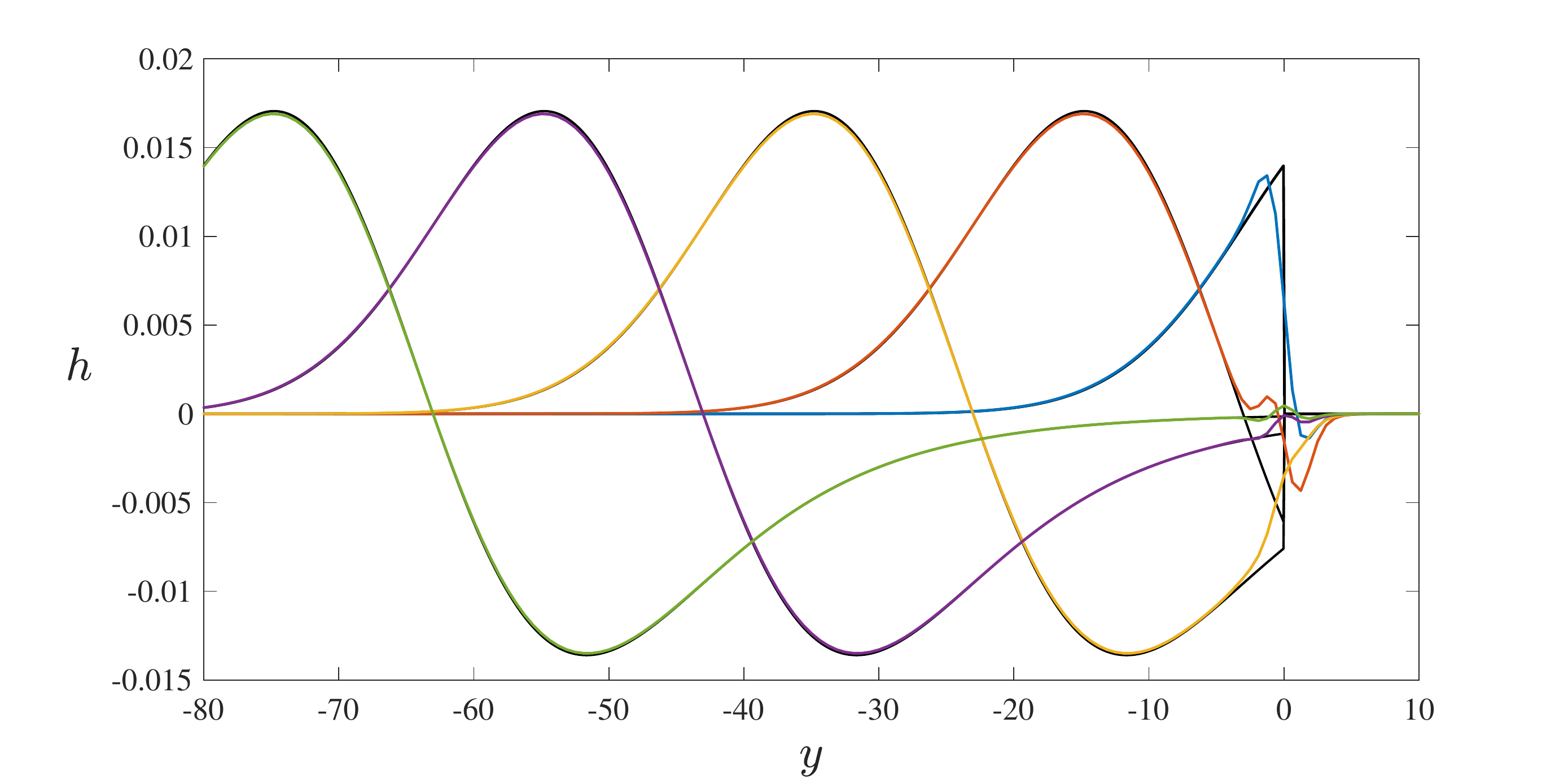}
\caption{Height field along the coast $x=0$ at times $t=-1$ (blue), $1$ (red), $3$ (landfall time, orange), $5$ (purple) and $7$ (green) for the simulation in figure \ref{fig:hAllEastward}. The predictions based on the matched asymptotics result \eqn{hc} are shown by the black curves.}
\label{fig:hCoastEastward}
\end{center}
\end{figure}

Figures \ref{fig:hAllEastward} and \ref{fig:hCoastEastward} are the analogues of figures \ref{fig:hAllWestward} and \ref{fig:hCoastWestward} for an eastward-travelling atmospheric perturbation coming from overland to reach the coast at $t=0$. 
In this case, the Kelvin-wave generation, visible at $t=-1$ in figure \ref{fig:hAllEastward} precedes the bulk of the QG response which appears only around $t=3$, when the perturbation is well away from the coast. As expected from \eqn{ms0} and \eqn{hc}, the Kelvin wave leads to an initial elevation of the sea surface, followed by a depression, reversing the evolution compared with the 
westward-travelling case. Figure \ref{fig:hCoastEastward} shows that the Kelvin-wave amplitude is again predicted with high accuracy by the asymptotic formula \eqn{hc}. 


\section{Discussion}

This paper discusses a simple example of spontaneous generation of  a Kelvin wave by a forced QG flow, thus demonstrating a fundamental limitation of the concept of balance in the presence of a boundary. The mechanism of wave generation is similar to the Lighthill radiation of acoustic waves by vortical flow: sufficiently long Kelvin waves are slow enough for their frequency to match that of the geostrophic flow, leading to a resonant response that is small, here $O(\eps)$, because of the mismatch between the spatial scales of the waves and geostrophic flow. The mechanism is robust and operative in initial-value problems as well as in forced problems such as the one considered here. The Kelvin-wave response to an unforced, initially well-balanced flow can in fact be obtained from the results of \citet{rezn-grim} on geostrophic adjustment in the presence of a boundary by requiring the initial conditions to be free of inertia-gravity and Kelvin waves. The only restriction to spontaneous Kelvin-wave generation is that the domain be large enough to allow for the propagation of Kelvin waves with $O(\eps^{-1} L_D)$ wavelengths.  

The paper focuses on the shallow-water model, but it is clear that the mechanism discussed applies to continuously stratified models as well, and that long baroclinic Kelvin waves can be  generated spontaneously by balanced motion. A non-trivial vertical structure offers an additional possibility of frequency matching, since Kelvin waves with $O(1)$ along-shore scales and $O(\eps)$ vertical scales 
are also slow (recall that the frequency of baroclinic Kelvin waves is proportional to the ratio of vertical to along-shore scales). 
These vertically-short Kelvin waves are however exponentially localised in an $O(\eps)$ boundary layer along the coast. As a result, they are only very weakly coupled to the interior QG flow, and their spontaneous generation can be expected to be exponentially small in $\eps$. To illustrate this point, we refer to the Kelvin-wave-induced instability of shear flows in a channel, which has been shown to have an exponentially small growth rate
\citep{v-yavn07}. However, our discussion ignores the steepening  and  shock formation that characterise the nonlinear dynamics of Kelvin waves \citep{rezn-grim,zeit18}. Vorticity generation by shocks provides a quite different mechanism of interaction between Kelvin waves and balanced flows,  examined in a baroclinic configuration by \citet{dewa-et-al}, \citet{dere-et-al} and \citet{vena20}. 

We conclude by emphasising the limitation of the standard QG model and its implicit assumption of $O(1)$ domain size. 
The filtering of Kelvin waves that the standard boundary condition  $h = C(t)$ entails is problematic for larger domains for two reasons. First, because of the lack of a frequency gap, balanced motion excites Kelvin waves with amplitudes that are algebraic in $\eps$; second, accounting for Kelvin waves is crucial to resolve the issue of non-conservation of the QG mass and boundary circulation \citep{rezn-suty}. This makes it desirable to obtain a version of the QG model that retains Kelvin waves (in the same way as the semi-geostrophic and L1 models do, see \citet{kush-et-al} and \cite{ren-shep}). At a linear level, this is straighforward: the boundary condition $h_y + \eps h_{xt} = \eps \Phi_x$, which approximates the exact condition \eqn{hbc} up to an $O(\eps^2)$ error, leads to a QG model that captures Kelvin waves and conserves mass exactly.  
Extensions to nonlinear dynamics and curved boundaries are worth considering. 
  
\medskip
\noindent
\textbf{Acknowledgments.} This work was supported by the UK Natural Environment Research Council grant NE/R006652/1. 


\bibliographystyle{unsrtnat}
\bibliography{mybib}

\begin{thebibliography}{20}
\providecommand{\natexlab}[1]{#1}
\providecommand{\url}[1]{\texttt{#1}}
\expandafter\ifx\csname urlstyle\endcsname\relax
  \providecommand{\doi}[1]{doi: #1}\else
  \providecommand{\doi}{doi: \begingroup \urlstyle{rm}\Url}\fi

\bibitem[Warn et~al.(1995)Warn, Bokhove, Shepherd, and Vallis]{warn-et-al}
T.~Warn, O.~Bokhove, T.~G. Shepherd, and G.~K. Vallis.
\newblock Rossby number expansions, slaving principles, and balance dynamics.
\newblock \emph{Quart. J. R. Met. Soc.}, 121:\penalty0 723--739, 1995.

\bibitem[Vanneste(2008)]{v08}
J.~Vanneste.
\newblock Exponential smallness of inertia-gravity-wave generation at small
  {R}ossby number.
\newblock \emph{J. Atmos. Sci.}, 65:\penalty0 1622--1637, 2008.

\bibitem[Vanneste(2013)]{v13}
J.~Vanneste.
\newblock Balance and spontaneous wave generation in geophysical flows.
\newblock \emph{Annu. Rev. Fluid Mech.}, 45:\penalty0 147--172, 2013.

\bibitem[Zeitlin(2018)]{zeit18}
V.~Zeitlin.
\newblock \emph{Geophysical fluid dynamics: understanding (almost) everything
  with rotating shallow water models}.
\newblock Oxford University Press, 2018.

\bibitem[Dorofeyev and Larichev(1992)]{doro-lari}
V.~L. Dorofeyev and V.~D. Larichev.
\newblock The exchange of fluid mass between quasi-geostrophic and ageostrophic
  motions during the reflection of rossby waves from a coast. {I}. the case of
  an infinite rectilinear coast.
\newblock \emph{Dynam. Atmos. Oceans}, 16:\penalty0 305--329, 1992.

\bibitem[Reznik and Grimshaw(2002)]{rezn-grim}
G.~M. Reznik and R.~Grimshaw.
\newblock Nonlinear geostrophic adjustment in the presence of a boundary.
\newblock \emph{J. Fluid Mech.}, 471:\penalty0 257--283, 2002.

\bibitem[Reznik and Sutyrin(2005)]{rezn-suty}
G.~M. Reznik and G.~G. Sutyrin.
\newblock Non-conservation of `geostrophic mass' in the presence of a long
  boundary and the related {K}elvin wave.
\newblock \emph{J. Fluid Mech.}, 527:\penalty0 235--264, 2005.

\bibitem[Kajiura(1962)]{kaji62}
K.~Kajiura.
\newblock A note on the generation of boundary waves of {K}elvin type.
\newblock \emph{J. Oceanogr. Soc. Japan}, 18:\penalty0 49--58, 1962.

\bibitem[Thomson(1970)]{thom70}
R.~E. Thomson.
\newblock On the generation of {K}elvin-type waves by atmospheric disturbances.
\newblock \emph{J. Fluid Mech.}, 42:\penalty0 657--670, 1970.

\bibitem[Gill and Schumann(1974)]{gill-schu}
A.~E. Gill and E.~H. Schumann.
\newblock The generation of long shelf waves by the wind.
\newblock \emph{J. Phys. Oceanogr.}, 4\penalty0 (1):\penalty0 83--90, 1974.

\bibitem[Grimshaw(1988)]{grim88b}
R.~Grimshaw.
\newblock Large-scale, low-frequency response on the continental shelf due to
  localized atmospheric forcing systems.
\newblock \emph{J. Phys. Oceanogr.}, 18:\penalty0 1906--1919, 1988.

\bibitem[Tang and Grimshaw(1995)]{tang-grim}
Y.-M. Tang and R.~Grimshaw.
\newblock A modal analysis of coastally trapped waves generated by tropical
  cyclones.
\newblock \emph{J. Phys. Oceanogr.}, 25:\penalty0 1577--1598, 1995.

\bibitem[Yankovsky(2009)]{yank09}
A.~E. Yankovsky.
\newblock Large-scale edge waves generated by hurricane landfall.
\newblock \emph{J. Geophys. Res.}, 114:\penalty0 C03014, 2009.

\bibitem[Vallis(2017)]{vall17}
G.~K. Vallis.
\newblock \emph{Atmospheric and oceanic fluid dynamics: fundamentals and
  large-scale circulation}.
\newblock Cambridge University Press, 2nd edition, 2017.

\bibitem[Vanneste and Yavneh(2007)]{v-yavn07}
J.~Vanneste and I.~Yavneh.
\newblock Unbalanced instabilities of rapidly rotating stratified shear flows.
\newblock \emph{J. Fluid Mech.}, 584:\penalty0 373--396, 2007.

\bibitem[Dewar et~al.(2011)Dewar, Berloff, and Hogg]{dewa-et-al}
W.~K. Dewar, P.~Berloff, and A.~{McC.} Hogg.
\newblock Submesoscale generation by boundaries.
\newblock \emph{J. Mar. Res.}, 69:\penalty0 501--522, 2011.

\bibitem[Deremble et~al.(2017)Deremble, Johnson, and Dewar]{dere-et-al}
B.~Deremble, E.~R. Johnson, and W.~K. Dewar.
\newblock A coupled model of interior balanced and boundary flow.
\newblock \emph{Ocean Modelling}, 119:\penalty0 1--12, 2017.

\bibitem[Venaille(2020)]{vena20}
A.~Venaille.
\newblock Quasi-geostrophy against the wall.
\newblock \emph{J. Fluid Mech.}, 894:\penalty0 R1, 2020.

\bibitem[Kushner et~al.(1998)Kushner, McIntyre, and Shepherd]{kush-et-al}
P.~J. Kushner, M.~E. McIntyre, and T.~G. Shepherd.
\newblock Coupled {K}elvin wave and mirage-wave instabilities in
  semi-geostrophic dynamics.
\newblock \emph{J. Phys. Oceanogr.}, 28:\penalty0 513--518, 1998.

\bibitem[Ren and Shepherd(1997)]{ren-shep}
S.~Ren and T.~G. Shepherd.
\newblock Lateral boundary contributions to wave-activity invariants and
  nonlinear stability theorems for balanced dynamics.
\newblock \emph{J. Fluid Mech.}, 345:\penalty0 287--305, 1997.

\end{thebibliography}

\end{document}